\documentclass[fleqn,usenatbib]{mnras}

\usepackage[T1]{fontenc}

\DeclareRobustCommand{\VAN}[3]{#2}
\let\VANthebibliography\thebibliography
\def\thebibliography{\DeclareRobustCommand{\VAN}[3]{##3}\VANthebibliography}

\usepackage{graphicx}	
\usepackage{amsmath}	
\usepackage{amssymb}	
\usepackage{newtxtext,newtxmath,multirow}

\newcommand{\psr}{PSR\,J0250$+$5854}
\newcommand{\xmm}{\textit{XMM--Newton}}

\title[X-ray nondetection of \psr{}]{X-ray nondetection of \psr{}}

\author[C. M. Tan et al.]{
C. M. Tan,$^{1,2}$\thanks{E-mail: \href{mailto:chia.tan@mcgill.ca}{chia.tan@mcgill.ca}}
M. Rigoselli,$^{3}$
P. Esposito,$^{4,3}$
B. W. Stappers$^{5}$
\\
$^{1}$Department of Physics, McGill University, 3600 rue University, Montr\'eal, QC H3A 2T8, Canada\\
$^{2}$Trottier Space Institute at McGill, McGill University, 3550 rue University, Montr\'eal, QC H3A 2A7, Canada\\
$^{3}$ INAF, Istituto di Astrofisica Spaziale e Fisica Cosmica Milano, via A.\ Corti 12, I-20133 Milano, Italy\\
$^{4}$ Scuola Universitaria Superiore IUSS Pavia, Palazzo del Broletto, piazza della Vittoria 15, 27100 Pavia, Italy\\
$^{5}$ Jodrell Bank Centre for Astrophysics, Department of Physics and Astronomy, The University of Manchester, Manchester M13 9PL, UK\\
}

\date{Accepted XXX. Received YYY; in original form ZZZ}

\pubyear{2022}

\begin{document}
\label{firstpage}
\pagerange{\pageref{firstpage}--\pageref{lastpage}}
\maketitle

\begin{abstract}

We conducted a deep \xmm{} observing campaign on the 23.5-s radio pulsar \psr{} in order to better understand the connection between long-period, radio-emitting neutron stars and their high-energy-emitting counterparts. No X-ray emission was detected resulting in an upper limit in the bolometric luminosity of \psr{} of $<$10$^{31}$ erg s$^{-1}$ for an assumed blackbody with a temperature of 85 eV, typical of an X-ray Dim Isolated Neutron Star (XDINS). We compared the upper limit in the bolometric luminosity of \psr{} with the known population of XDINSs and found that the upper limit is lower than the bolometric luminosity of all but one XDINS. We also compared \psr{} with SGR 0418+5729, the magnetar with low dipole magnetic field strength, where the upper limit suggests that if \psr{} has a thermal hot spot like SGR 0418+5729, it would have a blackbody temperature of $<$200\,eV, compared to 320\,eV of the magnetar.

\end{abstract}




\section{Introduction}

\psr{} is a radio pulsar discovered by the LOFAR Tied-Array All-Sky survey \citep{2018ApJ...866...54T}. 
It is one of the slowest-spinning radio pulsars known to date, with a rotational period of 23.5\,s. Very few other radio pulsars are known to have periods of more than 10\,s, such as PSR\,J2251$-$3711 ($P = 12.1$ s; \citealt{2020MNRAS.493.1165M}), PSR\,J0901$-$4046 ($P = 76$ s; \citealt{2022NatAs...6..828C}) and GLEAM-X\,J162759.5$-$523504.3~\citep{2022Natur.601..526H}, whose nature is currently debated but was purported to have a spin period of 18 minutes, but without a spin period derivative measurement.

\psr{} has a measured spin derivative of $2.7 \times 10^{-14}$ s s$^{-1}$. Assuming the spin-down of the pulsar is purely due to magnetic dipole braking, this suggests a characteristic age of $1.3 \times 10^{7}$ years and a high inferred surface dipole magnetic field of $2.6 \times 10^{13}$ G, compared to that of the typical radio pulsars of $\sim$10$^{12}$ G. \psr{} also has one of the lowest rotational energy loss rates at $8 \times 10^{28}$ erg s$^{-1}$. Using the YMW16 model, the dispersion measure of the pulsar places it at a distance of 1.6 kpc~\citep{2017ApJ...835...29Y}.

The rotational properties of \psr{} are found to be similar to the group of nearby, isolated neutron star that are detected in the soft X-ray band with no corresponding radio emission. Collectively known as the X-ray Dim Isolated Neutron Stars~\citep[XDINSs,][]{2008AIPC..983..331K,2009ASSL..357..141T}, they are characterised by a soft, blackbody-like continuum X-ray
emission, with temperatures ranging from 50--110\,eV, with no hard, non-thermal X-ray emission. However, recent analyses of the 0.2--10 keV spectrum of the XDINSs has shown that, at least two of them, have a non-thermal contribution \citep[see][and references therein]{2022MNRAS.516.4932D}. 
The rotation periods of XDINS lie between 3--17\,s and they have period derivatives of the order of $10^{-14}$\,s\,s$^{-1}$. This corresponds to rotational energy loss rates far lower than their X-ray luminosities, which are in the range of 10$^{31-32}$ erg s$^{-1}$.
The pulsed X-ray emission of XDINSs, with pulsed fractions between 1.5 and 18 percent, are likely to originate from a non-uniform temperature distribution on the stellar surface (see e.g. \citealt{2022MNRAS.513.3113R}). More recently, a radio pulsar PSR\,J0726$-$2612 was found to show X-ray emission similar to those of the XDINSs~\citep{2019A&A...627A..69R}, suggesting that the lack of radio emission from the other XDINSs could be due to unfavourable viewing geometry.

Another group of slowly-rotating neutron stars that often show thermal components in their spectra are the magnetars \citep{2015SSRv..191..315M,2015RPPh...78k6901T,2017ARA26A..55..261K,2021ASSL..461...97E}. They have periods between 1 and 12 s, 
and high quiescent X-ray luminosities (10$^{33-35}$ erg s$^{-1}$), generally larger than their loss rate of rotational energy. It is believed that they are powered by the decay of their ultra-high magnetic field \citep{1995MNRAS.275..255T,1996ApJ...473..322T}.
Usually, magnetars are rather young objects (ages $\sim$10$^3$--10$^4$ years), as inferred from their timing parameters and/or from supernova remnant associations, but a few of them have period derivatives smaller than $10^{-13}$ s~s$^{-1}$ and thus characteristic ages larger than 10$^6$ years and magnetic fields of the order of $10^{13}$ G. They are SGR\,0418+5729 \citep{2010Sci...330..944R,2013ApJ...770...65R}, Swift\,J1822.3$-$1606 \citep{2012ApJ...761...66S,2014ApJ...786...62S}, and 3XMM\,J1852+0033 \citep{2014ApJ...781L..17R}.
The small number of detected high-energy bursts from these objects and their low quiescent X-ray luminosity (10$^{30}$--10$^{31}$ erg s$^{-1}$) have led to the hypothesis that these might be worn-out magnetars, approaching the end of their active life \citep{2011ApJ...740..105T,2013MNRAS.434..123V}.

\psr{} was previously observed with the \textit{Neil Gehrels Swift Observatory’s} X-ray Telescope~\citep[\textit{Swift}/XRT,][]{2005SSRv..120..165B} for two epochs in 2018 March, for a total exposure time of 10\,ks. There was no X-ray emission detected from the pulsar, and only a weak constraint in the bolometric luminosity of the pulsar at $<$$2 \times 10^{33}$ erg s$^{-1}$ \citep{2018ApJ...866...54T} could be placed. 
The constraint is an order of magnitude higher than the bolometric luminosity of the brightest XDINS, RX\,J0720.4$-$3125.

In order to understand better the link between \psr{} and the high-energy-emitting neutron stars, we requested a deep \xmm{} campaign to observe the pulsar. Section~\ref{sec:obs} describes the observations conducted by \xmm{} and the data reduction process. Section~\ref{sec:analysis} describes the analysis of the observations and contains the discussion from the obtained results. Section~\ref{sec:conclusion} provides a summary of the results regarding the link between \psr{} and the high-energy-emitting neutron stars.

\section{Observation and Data Reduction} \label{sec:obs}

\psr{} was observed by \xmm{} on 11 different epochs, with 10 epochs between 2019 July 28 and 2019 September 17, and the latest on 2020 August 28, yielding a 
total exposure time of 108\,ks. The observations were conducted using the three European Photon Imaging Camera (EPIC) instruments on board the spacecraft, which are sensitive in the 0.15--15 keV energy band. The instruments were operated in full frame mode. Due to the presence of several bright optical sources nearby, the medium optical-blocking filters were used for the two EPIC-MOS detectors and the thin filter was used for the EPIC-pn detector.

The data were processed using the \xmm{} Science Analysis Software tools (SAS). The EPIC-pn and EPIC-MOS data were processed with the \textsc{epproc} and \textsc{emproc}, respectively. In order to maximise the sensitivity at low energies ($<$0.25 keV), the algorithm \textsc{epreject} was used to reduce the background component, mainly due to detector noise, on processing the EPIC-pn observations. We filtered out time intervals with high background and applied the standard pattern selection criteria (patterns 0--4 for EPIC-pn and 0--12 for EPIC-MOS). 
This resulted in a combined net exposure time of 50.3\,ks for the EPIC-pn observation, as well as 94.5\,ks and 83.5\,ks for the two EPIC-MOS camera, respectively.

\begin{figure}
    \centering
    \includegraphics[width=\linewidth]{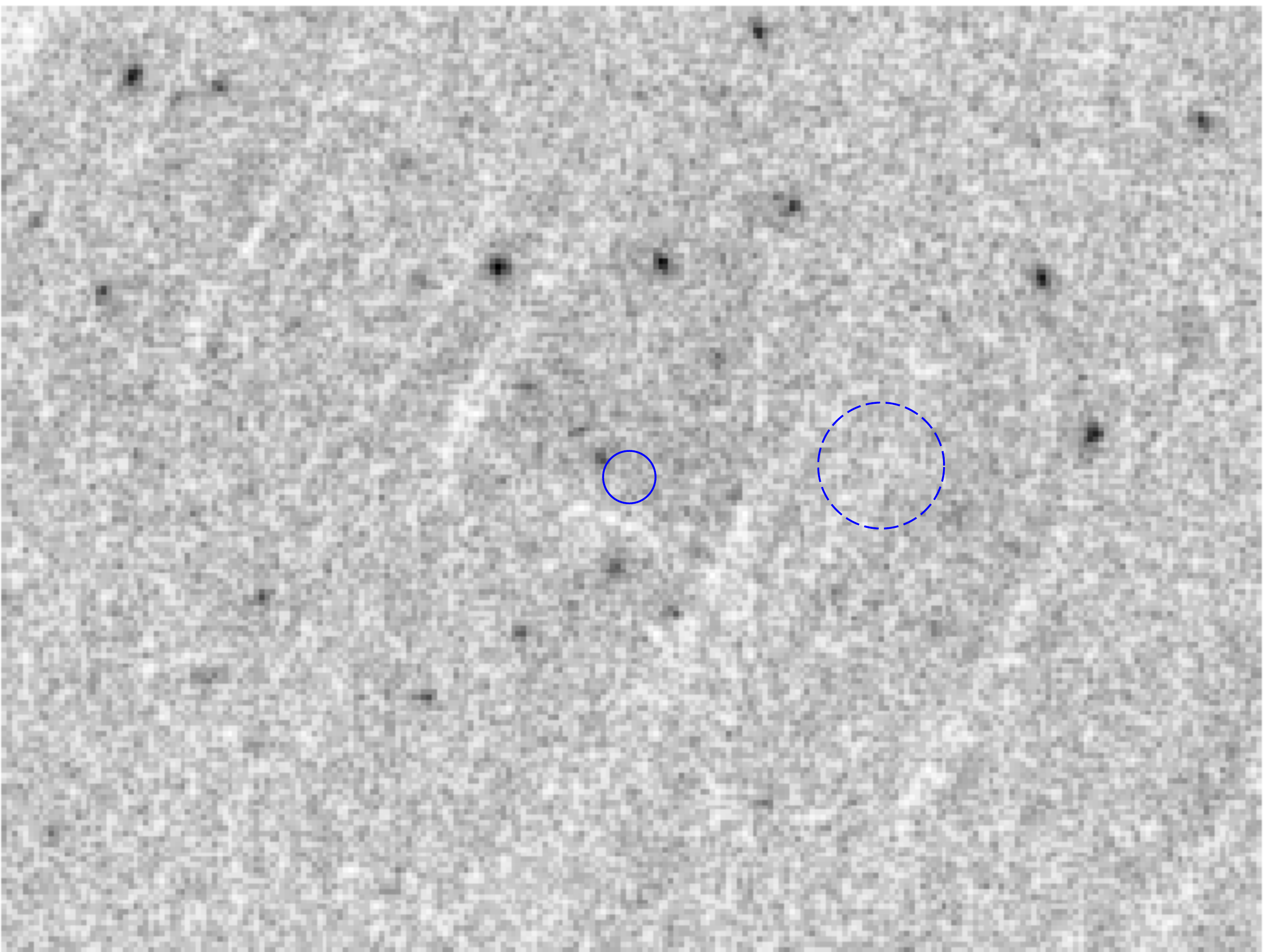}
    \caption{A $20' \times 15'$ EPIC-pn image at energies between 0.3--2 keV around the location of \psr{}, formed by combining all the data from individual observations. The solid circle indicates a $25"$ radius around the radio timing position of the pulsar. The dashed circle indicates a nearby background region of $60"$ radius.}
    \label{fig:pn_image}
\end{figure}

The filtered event lists generated from individual observations of each detector were then merged using the SAS tool \textsc{merge} to form combined event lists. The combined event lists for each instrument were then searched for X-ray emission from \psr{}. No X-ray sources are detected within $25"$ of the expected position of the pulsar in any of the three instruments, taking into account the systemic uncertainty of the EPIC-MOS ($15"$) and EPIC-pn ($4"$) cameras, respectively. Figure~\ref{fig:pn_image} shows the combined image from the EPIC-pn camera, centred around the best known position of the pulsar.

The SAS tool \textsc{eupper} was then used to estimate the 3$\sigma$ upper limit on the X-ray count rate from \psr{}, using a source extraction region of $25"$ around the radio position of the pulsar and a nearby background region of size $60"$, as indicated in Figure~\ref{fig:pn_image}. An upper limit is derived from the EPIC-pn data, while a single upper limit is derived from the two EPIC-MOS cameras. The 3$\sigma$ upper limit in count rate of \psr{} in the energy band of 0.3--2 keV is shown in Table~\ref{tab:upper_limit}.

\begin{table}
	\centering
	\caption{The net exposure time and 3$\sigma$ upper limit on the count rate from the nondetection of \psr{} in the energy band of 0.3$-$2 keV, for both the EPIC-pn and EPIC-MOS (combined) cameras.}
	\label{tab:upper_limit}
	\begin{tabular}{ccc} 
		\hline
		Instrument & Net exposure time & Upper limit\\
		keV & ks & count/s\\
		\hline
		pn & 50.3 & $9.0 \times 10^{-4}$ \\
	    MOS1 & 94.5 & \multirow{2}{*}{$8.2 \times 10^{-4}$} \\
	    MOS2 & 83.5 &  \\
		\hline
	\end{tabular}
\end{table}

\section{Analysis and Discussion} \label{sec:analysis}

The upper limit in count rates from the X-ray nondetection of \psr{} is used to compute the upper limit in bolometric luminosity for an XDINS-like object. Two spectra are extracted from the EPIC-pn and EPIC-MOS observations, which are used by the XSPEC software to convert the upper limits in count rates to upper limits in the unabsorbed X-ray fluxes and subsequently bolometric luminosity. We computed the upper limits for blackbody temperatures of 50, 85 and 115 eV, which are within the range of temperatures of the known XDINSs~\citep{2013MNRAS.434..123V}. An estimate of the neutral hydrogen column density in the direction of \psr{} is made using the dispersion measure of the pulsar and the usual assumption of 10 percent ionized interstellar medium  \citep{2013ApJ...768...64H}, which gives a value of $N_{\rm H} = 1.36 \times 10^{21}$ cm$^{-2}$. The upper limit on the bolometric luminosity is computed assuming the distance to the pulsar of 1.6 kpc, based on the dispersion measure-distance model of \citet{2017ApJ...835...29Y}. The computed upper limits are shown in Table~\ref{tab:ul_flux_lum}.

\begin{table}
	\centering
	\caption{The upper limits on the unabsorbed flux and bolometric luminosity of \psr{} for an energy range of 0.05$-$10 keV, derived from the EPIC-pn and 0.01$-$10 keV from the EPIC-MOS cameras. The bolometric luminosity was computed assuming $d=1.6$ kpc.}
	\label{tab:ul_flux_lum}
	\begin{tabular}{cccc} 
		\hline
		Instrument & Temperature & Upper limits & Bolometric luminosity\\
		& & flux & \\
		 & (eV) & (erg cm$^{-2}$ s$^{-1}$) & (erg s$^{-1}$)\\
		\hline
		pn & 115 & $7.0 \times 10^{-15}$ & $2.0 \times 10^{30}$\\
		 & 85 & $1.6 \times 10^{-14}$ & $4.5 \times 10^{30}$\\
		 & 50 & $1.7 \times 10^{-13}$ & $5.0 \times 10^{31}$\\
	    MOS & 115 & $2.2 \times 10^{-14}$ & $6.5 \times 10^{30}$\\
	     & 85 & $4.5 \times 10^{-14}$ & $1.3 \times 10^{31}$\\
	     & 50 & $3.9 \times 10^{-13}$ & $1.1 \times 10^{32}$\\
		\hline
	\end{tabular}
\end{table}

\begin{figure}
    \centering
    \includegraphics[width=\linewidth]{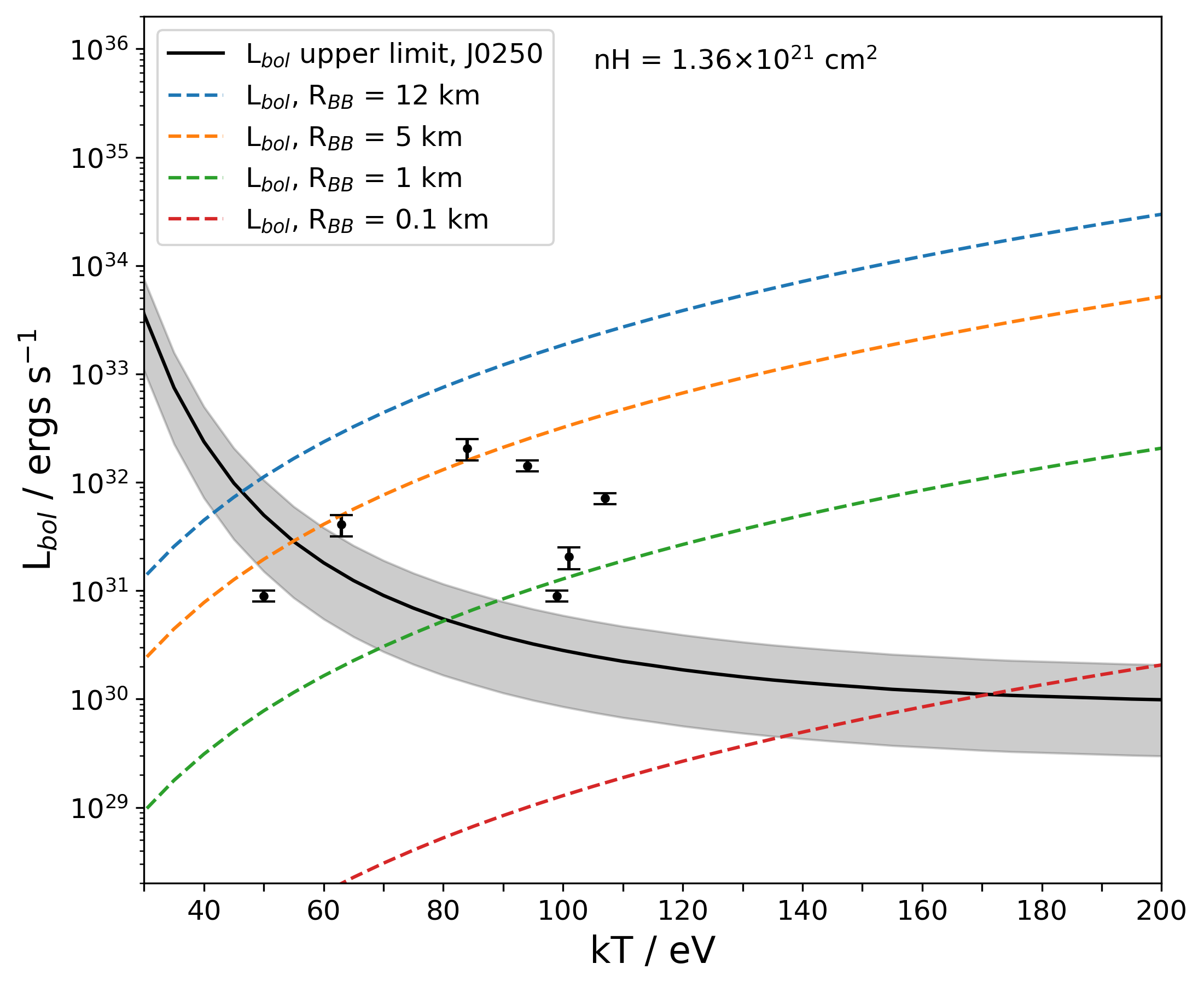}
    \caption{The upper limit in the bolometric luminosity of \psr{} from the EPIC-pn data, for various values of blackbody temperature between 30--200 eV. The shaded region is the uncertainty in the computed upper limit, considering $d=1.6\pm0.7$ kpc as explained in the text. The dashed lines are the expected bolometric luminosity of a blackbody of size 12, 5, 1, and 0.1 km, respectively. The black dots refer to the 7 known XDINSs, with data taken from \citet{2013MNRAS.434..123V}.}
    \label{fig:kT_Lbol}
\end{figure}

As the EPIC-pn data provided a lower upper limit in bolometric luminosity for all the trialled temperatures, we only used this data set for further interpretation of the X-ray nondetection of \psr{}. We compute the upper limits in the bolometric luminosity of \psr{} at various blackbody temperatures between 30--200 eV, using the same assumption on the neutral hydrogen column density and distance to the pulsar. The upper limits are plotted as a black solid line in Figure~\ref{fig:kT_Lbol}. We estimated an uncertainty on the distance to the pulsar of 45 per cent, considering the assertion from \citet{2017ApJ...835...29Y}, that 95 per cent of the distance estimates have a relative error of less than a factor of 0.9. This resulted in an uncertainty in the computed upper limit, which is plotted as a grey shaded region in Figure~\ref{fig:kT_Lbol}.

We then compared these upper limits for the luminosity with values computed from different manifestations of thermal-emitting cooling neutron stars. The thermal emission can arise from the whole surface of the star (typical radius of 12 km), or from a some fraction of the surface. 
We computed the bolometric luminosity for temperatures between 30--200 eV and radii of 12 km (dashed blue line in Figure~\ref{fig:kT_Lbol}), 5 km (dashed orange line) and 1 km (dashed green line). Using the most conservative estimate on the distance to the pulsar, we found that the upper limits on \psr{} emission are consistent with temperatures lower than 50, 60 and 90 eV, respectively, and bolometric luminosities lower than 10, 3.5 and 0.9 $\times 10^{31}$ erg s$^{-1}$. As a comparison, we added the bolometric luminosities of the seven XDINSs with black dots \citep{2013MNRAS.434..123V}. They are compatible with neutron stars with a blackbody emitting radius between 1 and 5 km and, with the exception of RX J0420.0$-$5022, they all exceed the upper limit we infer for \psr{}.
We caution that these limits are derived with the assumption of purely blackbody emission with a single temperature, while it is well known that XDINSs spectra are modelled by two distinct blackbodies with different temperature and size, and often containing one or more Gaussian absorption features \citep{2019PASJ...71...17Y}.

The derived upper limits suggest that, if \psr{} is indeed an XDINS-like object, the pulsar is cooler than most known XDINSs. This would not be a surprise, considering the characteristic age of \psr{} of $\sim$10$^{7}$ years is higher than those of XDINSs (10$^{5}$--10$^{6}$ years). Based on the neutron star cooling model of \citet{2013MNRAS.434..123V}, the upper limit in the bolometric luminosity of \psr{} would yield a lower limit on cooling age of the neutron star of $2 \times 10^{6}$ years, lower than its characteristic age.


If \psr{} is instead a low-$B$ old magnetar, according to the evolutionary tracks in the $P$--$\dot P$ diagram of \citet{2013MNRAS.434..123V}, it should be a descendant of magnetars with extremely high magnetic fields ($\sim$10$^{15}$ G). It should have a real age of a few $10^5$ years and a bolometric luminosity of $\sim$10$^{33}$ erg s$^{-1}$, that is ruled out by our measurement for any $kT\gtrsim35$ eV.
However, the low-$B$ magnetar SGR 0418+5729 has a similar true age and X-ray luminosity $L_{\mathrm{X}} \approx 10^{31}$~erg s$^{-1}$, and the bulk of its thermal luminosity comes from a small hot spot of 320 eV \citep{2013ApJ...770...65R}, which is not related to long-term cooling but rather is maintained by the bombardment of energetic particles carried by magnetospheric currents \citep{2015MNRAS.452.3357G}.
To test such a scenario, we computed the bolometric luminosity for temperatures between 30--200 eV and a radius of 0.1 km (dashed red line in Figure~\ref{fig:kT_Lbol}). The upper limit to the blackbody temperature and bolometric luminosity, using the conservative distance estimate, are $200$ eV and $2.0 \times 10^{30}$ erg s$^{-1}$, respectively.

\section{Conclusions} ~\label{sec:conclusion}

The upper limits on the bolometric luminosity of \psr{} derived from the \xmm{} observations suggest that the pulsar is unlikely to be an XDINS, unless it has an effective blackbody temperature of $<$50 eV, placing it at the very low end of the known distribution of XDINSs. 
Our conclusion holds regardless of the specific value of the distance and its uncertainty; indeed, the upper limit of \psr\ bolometric luminosity would match the XDINS ones if increased by a factor $\sim$50, that would imply a distance of about 10 kpc. This value seems rather unlikely, given that our pulsar has a dispersion measure DM $=45.28$ pc$^{-3}$, while the max DM along the line-of-sight is 295.5 pc cm$^{-3}$ \citep{k22}.

\psr{} could also be a low-$B$ old magnetar with a heated hot spot, similar to SGR 0418+5729. However this would only be the case if the temperature of the hot spot is at least 1.6 times smaller, at $<$200\,eV. The results indicate that if \psr{} is related to the high-energy-emitting neutron stars, the object would have been much cooler than most of the known XDINSs and magnetars.

\section*{Acknowledgements}

Based on observations obtained with \xmm, an ESA science mission with instruments and contributions directly funded by ESA Member States and NASA. MR acknowledge financial support from the Italian Ministry for Education, University and Research through grant 2017LJ39LM `UnIAM' and the INAF `Main-streams' funding grant (DP n.43/18). We thank Jason Hessels for his contribution to the discussion of the work.

\section*{Data Availability}
The X-ray data are available through the \xmm Science Archive at \href{}{https://www.cosmos.esa.int/web/xmm-newton/xsa}. Obs.IDs: 0844000201, 0844000301, 0844000401, 0844000501, 0844000601, 0844000701, 0844000801, 0844000901, 08440001001, 0844001101, and 0844001201. 
 



\bibliographystyle{mnras}
\bibliography{biblio} 








\bsp	
\label{lastpage}
\end{document}